# Web 2.0 Technologies and Social Networking Security Fears in Enterprises

Fernando Almeida
INESC TEC, Faculty of Engineering, University of Porto
Rua Dr. Roberto Frias, 378, 4200-465 Porto, Portugal

*Abstract*— **Web 2.0 systems have drawn the attention of corporation, many of which now seek to adopt Web 2.0 technologies and transfer its benefits to their organizations. However, with the number of different social networking platforms appearing, privacy and security continuously has to be taken into account and looked at from different perspectives. This paper presents the most common security risks faced by the major Web 2.0 applications. Additionally, it introduces the most relevant paths and best practices to avoid these identified security risks in a corporate environment.**

## I. INTRODUCTION

Applications are the lifeblood of today's organization as they allow employers to perform crucial business tasks. When granted access to enterprise networks and the Internet, applications can enable sharing of information within workgroups, throughout an enterprise and externally with partners and customers. Until recent years, when applications were launched only from desktop computers and servers inside the corporate network, data security policies were relatively easy to enforce. However, today's organizations are grappling with a new generation of security threats. Consumer-driven technology has unleashed a new wave of Internet-based applications that can easily penetrate and circumvent traditional network security barriers.

The Web 2.0 introduces the idea of a Web as a platform. The concept was such that instead of thinking of the Web as a place where browsers viewed data through small windows on the readers' screens, the Web was actually the platform that allowed people to get things done. Currently this initial concept has gained a new dimension and is really starting to mean a combination of the technology allowing customers to interact with the information [1].

Social-networking Web sites, such as Facebook and MySpace, now attract more than 100 million visitors a month [2]. As the popularity of Web 2.0 has grown, companies have noted the intense consumer engagement and creativity surrounding these technologies. Many organizations, keen to harness Web 2.0 internally, are experimenting with the tools or deploying them on a trial basis.

Reference [3] admits that Web 2.0 could have a more far-reaching organizational impact than technologies adopted in the 1990s (e.g., enterprise resource planning (ERP), customer relationship management (CRM), and the supply chain management (SCM)). The organizational of these new collaborative platforms are illustrated in figure 1. The latest Web tools have a strong bottom-up element and engage a broad base of workers. They also demand a mind-set different from that of earlier IT programs, which were instituted primarily by edicts from senior managers.

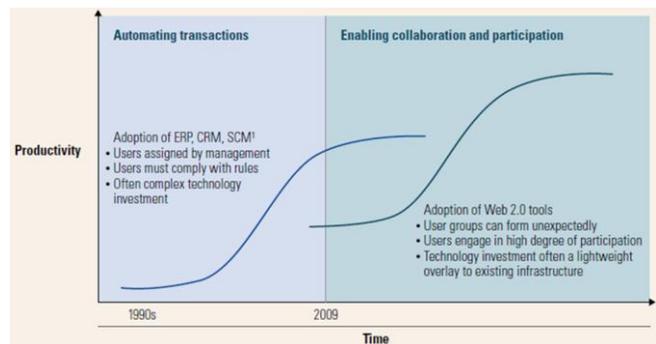

Figure 1. Adoption of corporate technologies. [4]

Web 2.0 covers a wide range of technologies. The most widely used are blogs, wikis, podcasts, information tagging, prediction markets, and social networks. A short description of these technologies potentialities is given in table 1.

TABLE 1. WEB 2.0 TECHNOLOGIES [4]

| Web 2.0 technologies | Description | Category of technology |
|---|---|---|
| Wikis, shared workspaces | Facilitates co-creation of contents across large and distributed set of participants. | Broad collaboration. |
| Blogs, podcasts, videocasts | Offers individuals a way to communicate and share information with other people. | Broad communication. |
| Prediction markets, polling | Harnesses the power of community and generates a collectively derived answer. | Collective estimation. |
| Tagging, user tracking, ratings, RSS | Add additional information to primary content to prioritize information. | Metadata creation. |
| Social networking, network mapping | Leverages connections between people to offer new applications | Social graphing. |

New technologies are constantly appearing as the Internet continues to evolve. What distinguishes them from previous technologies is the high degree of participation they require to be effective [5]. Unlike ERP and CRM, where most users either simply process information in the form of reports or use the technology to execute transactions (such as issuing payments or entering customer orders), Web 2.0 technologies are interactive and require users to generate new information and content or to edit the work of other participants.

These new Internet-based communications tools such as Facebook, Twitter and Skype have already achieved widespread penetration inside organizations [6]. Inevitably, these new Internet-based technologies and applications have spawned a new set of challenges for enterprises seeking to secure their networks against malicious threats and data loss. Allowing employees to access Web 2.0 applications has made enforcing data security policies a far more complex problem. Even worse, many businesses have no way to detect, much less control these new applications, increasing the potential for intentional or accidental misappropriating of confidential information.

## II. WEB 2.0 ADOPTION IN ORGANIZATIONS

Web 2.0 solutions are used for a variety of business purposes. According to survey study conducted by Gartner [6], about half of the organizations employ Web 2.0 solutions for IT functions, and roughly a third of organizations use them for marketing, sales or customer service. One in five organizations reported using Web 2.0 for public relations or human resources, particularly in the recruitment field [7]. The same study also establishes that by 2014, social networking services will replace e-mail as the primary vehicle for interpersonal communications, for 20 percent of the business users.

Another study conducted by [8] on the end of 2010 reports that Web 2.0 continues to grow, showing significant increases in the percentage of companies using social networking (40 percent) and blogs (38 percent). Furthermore, this survey shows that the number of employees using the dozen Web 2.0 technologies continues to increase. On the same way, nearly two-thirds of respondents at companies using Web 2.0 say they will increase future investments in these technologies, compared with just over half in 2009 [8].

The most common business benefits from using Web 2.0 based on the literature revision includes the increasing speed of access to knowledge, reducing communication costs, increasing effectiveness of marketing, increasing customer satisfaction, increase brand reputation, increasing speed of access to knowledge and reducing communication costs [9] [10]. Different types of networked organizations can achieve different benefits, namely:

- Internally networked organizations – some companies are achieving benefits from using Web 2.0 primarily within their own corporate walls. In this case, Web 2.0 is integrated tightly into their workflows and promotes significantly more flexible processes;
- Externally networked organizations – other companies achieve substantial benefits from interactions that spread beyond corporate borders by using Web 2.0 technologies to interact customers and business partners;
- Fully networked organizations – finally, some companies use Web 2.0 in revolutionary ways. They derive very high levels of benefits from Web 2.0's widespread use, involving employees, customers, and business partners.

## III. WEB 2.0 SECURITY RISKS

The collaborative, interactive nature of Web 2.0 has great appeal for business from a marketing and productivity point of view. Companies of all sizes and vertical markets are currently taking full advantage of social networking sites such as Facebook, Twitter and LinkedIn to connect with colleagues, peers and customers. In fact, not only these technologies are useful, but companies that don't adapt could well find themselves left behind the social revolution [11].

Companies are leveraging these sites for more than just communicating. Through Web 2.0 and social networking areas, enterprises are exchanging media, sharing documents, distributing and receiving resumes, developing and sharing custom applications, leveraging open source solutions, and providing forums for customers and partners [12].

While all this interactivity is exciting and motivating, there is an enterprise triple threat found in Web 2.0: losses in productivity, vulnerabilities to data leaks, and inherent increased security risks.

There are certain organizations that embrace Web 2.0 usage by employees, but the majority of them follow a different approach. Reference [6] shows that eighty-one percent of organizations restrict the use of at least one Web 2.0 tool because they are concerned about security. Therefore, organizations restrict social media usage through policy, technology and controlling the use of user-owned devices. While blocking access to social media provides better security, it is widely accepted that it is never feasible nor sustainable in the face of emerging use in the 21st century. Instead, we are living in a future where organizations must plan and design environments with less control of employee activities.

As referred previously in this paper, the primary concern that organizations have about employee usage of Web 2.0 technologies is security. Figure 2 illustrates the top concerns perceived by companies about the use of Web 2.0 technologies.

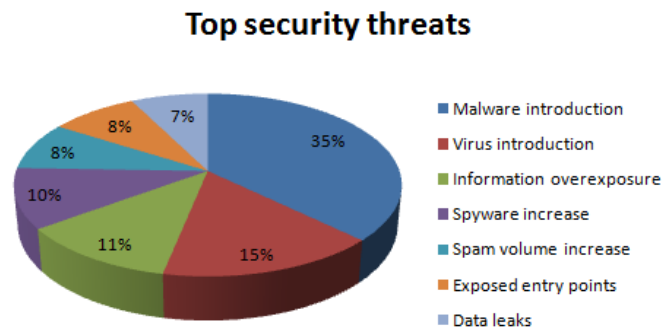

Figure 2.  Top security threats of Web 2.0 usage. [7]

The security concern is a specific obstacle to adoption and integration of social media in organizations. The top four perceived threats from employees' use of Web 2.0 are malicious software (35 percent), viruses (15 percent), overexposure of information (11 percent) and spyware (10 percent).

As with any evolution of a product or service, the old ways of performing a task or providing a solution simply may not work. This is also true in reducing and mitigating Web 2.0 threats. Time tested security solutions are no longer the key defense in guarding against attacks and data loss. Some characteristics of 2.0 securities that are being discussed in the literature are:

- Traditional Web filtering is no longer adequate;
- New protocols of AJAX, SAML and XML create problems for detection;
- RSS and rich Internet applications can enter directly into networks;
- Non-static Web content makes identification difficult;
- High bandwidth use can hinder availability;
- User-generated content is difficult to contain.

Security teams of a company must be aware of the need to address Web 2.0 threat in their desktop clients, protocols and transmissions, information sources and structures, and server weakness. In fact, none of these attack vectors are new, but the way to respond to them may be. Many of the threats are obvious associated with Internet use, but there are others, particularly the ones that can lead to confidential data loss, that can be addressed and mitigated by enterprises.

Direct posting of company data to Web 2.0 technologies and communities is the most common. No vulnerability need to be exploited or malicious code injected when employees (whether as part of their responsibilities or not) simply post protected or restricted information on blogs, wikis, or social networking sites. According to many security companies, the attacks on these technologies are on the rise. Many of these attacks also come via malicious payloads, which are downloaded when spam and phishing scams are utilized. According to Sophos [13], 57% (an increase of over 70% from the previous year) of people who use social networks report receiving spam and phishing messages.

Some security concerns are specific to Web 2.0 tools used by employees. For example, technologies that are perceived to facilitate work productivity, such as webmail, collaborative platforms and content sharing applications, are less likely to raise concern than the mainstream social media tools such as Facebook, LinkedIn, YouTube and Twitter, which are typically not allowed by companies.

There are both real and perceived consequences of inappropriate Web 2.0 and social media use:

- The financial consequence for security incidents (including downtime, information and revenue loss);
- The inappropriate use of social media may loss company reputation, brand or client confidence;
- Additional unplanned investments necessary for implementing social media in their organizations;
- Litigation of legal threats caused by employees disclosing confidential or sensitive information.

Organizational leaders are facing real consequences when adopting Web 2.0 technologies, but they recognize a growing demand for employee usage. The CIOs must find the delicate balance between security and the business need for these tools, and enable their use in such a way that reduces the risk for data loss or reputational harm to the corporate brand. While a sound security policy is a necessity in proactively responding to Web 2.0, policies must be enforced by technology.

### IV. PATHS AND BEST PRACTICES TO AVOID SECURITY RISKS

Most enterprises already have a form of an acceptable use policy, which should govern the use of all resources in the enterprise computing environment. The Web 2.0 application evaluation should form part of the organization's risk management process. Organizations can implement policies restricting employees' use of Web 2.0 applications. The following guidelines should be followed when formulating the policy:

- The policy should be created after consultation with all stakeholders;
- The policy should be based on principles, but should be detailed enough to be enforceable;
- The policy must be effectively communicated;
- Policies should be aligned with those already in operation relating to, for example, e-mails.

Some best practices should be taken into account when implementing the policy. First, responsibility of implementing and enforcing the policy should be shared and delegated to the various departments. Second, a compliance officer should be made accountable for the oversight and co-ordination function. Finally, all users must acknowledge the policy in writing. We must always consider that a policy is only effective if it is known and understood by all users.

The users should be aware and educated on the safeguards and policies. Therefore, users must be trained to identify Web 2.0 applications and understand the risks, as well as stay informed about the latest news on fraudulent Internet activities. Employees should understand and implement the security feature, which these websites provide.

Critically read the current policy in a context of 2.0 technologies, and identify gaps that need to be addresses, is a fundamental task for a CIO. For instance, because the risks and the inherent difficulty managing the use of social networking applications, many enterprises have made the decision to not allow access to social networking services and Web 2.0 powered sites from inside the corporate perimeter. Of greatest importance is a clear and unambiguous warning in the policy about sharing confidential corporate information.

Enforcement of the policy can be made through analysis of Web logs for use during business time, or through automated searches of websites for corporate information. According to Gartner [4], many organizations have already included Web 2.0 and data protection sections to their training on protecting corporate information.

In the following, we will present some IT policies that should be included to allow a safe inclusion of 2.0 technologies in the enterprise environment.

*A. Application control list*

Network traffic and applications are generally controlled at the firewall by tracking the ports used, source and destination addresses, and traffic volume. However, these methods may not be sufficient to precisely define or control the traffic from Internet-based applications. To address this problem, we must use protocol decoders to decrypt and examine network traffic for signatures unique to an application. In this way, even when applications attempt to hide by using non-standard ports and protocols, they can still be discovered. In addition, protocol decoders enable decryption and examination of encrypted network traffic. This allows application control to be applied to IPSec and SSL-encrypted VPN traffic, including HTTPS, POP3S, SMTPS and IMAPS protocols.

Applications that need to be explicitly managed are entered into an application control list in the firewall policy. Administrators can create multiple application control lists, each configured to allow, block, monitor or shape network traffic from a unique list of applications. An application "whitelist" is appropriate for use in a high security network as it allows only traffic from listed applications to pass through the gateway. On the other hand, an application "blacklist" allows all unlisted application traffic to pass. Applications can be controlled individually or separated into categories and controlled as groups.

*B. Application traffic shaping*

Application traffic shaping allows administrators to limit or guarantee the network bandwidth available to all applications or individual applications specified in an application list entry. For example, a business could limit the bandwidth used by Skype and Facebook chat to no more than 100 kilobytes per second, or restrict YouTube traffic to reserve network bandwidth for mission critical applications. Traffic shaping can also be configured on a time-sensitive basis to restrict user access or bandwidth available to applications during certain times of the day. Traffic shaping policies must be created independently of firewall policies and application control lists so that administrators can reuse them in multiple policies and list entries. Shared traffic shaping policies can be applied to individual firewalls or across all firewalls.

*C. Monitoring and review*

Extensive logs and audit trails should be maintained. These should be regularly reviewed, with a reporting system implemented. The application monitoring and reporting feature may collects application traffic information and displays it using visual trend charts, giving administrators a quick way to gain insight into application usage on their networks.

The most relevant trend charts (e.g., top blocked websites, top ten applications by bandwidth) may be generated for each firewall policy that has application monitoring enabled. Using the knowledge gained from application trend charts, administrator can quickly optimize the use of applications considering the organization's security policy and worker needs.

*D. Browser settings*

Browser and security settings should be customized to its highest level. Alternatively, non-standard browsers (such as Mozilla Firefox) that allow for anonymous surfing and which are equipped with advanced functions can be used. Users should note whether the security features (such as HTTPS) on the browser are operating.

*E. Anti-malware software*

Installing anti-virus, anti-spyware programs and spam-filters for both inbound and outbound traffic should be implemented at the gateway and desktop levels. Anti-malware software, with the following functionalities, should be implemented:

- Messages should be deep-scanned, searching for signature patterns, placing reliance heuristic or behavioral based protection [14];
- Virus scanners should be able to detect any treat and update the network and firewall rules immediately;
- Utilize software that decomposes all container file types its underlying parts, analyzing the underlying parts for embedded malware.

*F. Authentication*

Strong or non-password based authentication should be deployed and used for access to sensitive information and resources. Web 2.0 applications usually employ weak authentication, and are targets for a chain of penetration and social engineering attacks that can compromise valuable resources. Requiring appropriate token-based or biometric authentication at key points can help to prevent incidents.

*G. Avoid clickjacking*

Currently, clickjacking is considered one of the most dangerous and troubling security problems on the Web [15]. In this attack, two layers appear on a site, one visible, one transparent, and users inadvertently interact with the transparent layer that has malicious intent. New countermeasures, such as NoScript with ClearClick reduce the clickjacking risk. Additionally, users can take other countermeasures to limit clickjacking risk, such as minimizing cookie persistence by logging out of applications and using a dedicated browser for each website visited.

*H. Data loss protection*

Data infiltration is a continuing challenge of organizations participating in the Web 2.0 environment. Protecting the integrity and confidentiality of organizational information from theft and inadvertent loss is a key issue today. The cost of dealing with a data breach continues to rise each year. In September of 2011, Reference [16] conducted a study that reveals that the average cost to an enterprise from a data breach rose from $7.12 million in 2010 to $7.25 million in 2011.

The implementation of a data loss protection (DLP) solution may be integrated with 2.0 technologies. The DLP, based on central policies, will be responsible to identify, monitor and protect data at rest, in motion, and in use, through deep content analysis.

The DLP solutions both protect sensitive data and provide insight into the use of content within the enterprise. Few enterprises classify data beyond public vs. everything else [17]. Therefore, DLP helps organizations better understanding their data, and improves their ability to classify and manage.

## V. Conclusion

As we enter the second decade of the 21st century, the landscape of communication, information and organizational technologies continues to reflect emerging technological capabilities, as well as, changing user demands and needs. The Web 2.0 is a typical term used to describe these social technologies that influence the way people interact. Simultaneously, these Web 2.0 technologies are coming to the enterprise, radically transforming the way employees, customers and applications communicate and collaborate.

These technological advancements will continue to bring new opportunities and threats, thus requiring agility and continued evolution of resources. Successful organizations will be those that determine where and how to embrace these emergent tools to add new value and agility to their organizations. Success will require careful, on-going efforts to safeguard assets, including infrastructure, data, and employees, along with measured and educated adoption of new cyber technologies.

A comprehensive security program should be adopted by companies to deal with the introduction of Web 2.0 technologies in a corporate environment. As a first step, a Web 2.0 policy should be formulated implemented and the compliance with the policy should be monitored. The policy should be easy to understand, implemented and monitored, yet, detailed enough to be enforceable and be used to hold users accountable. Users should be trained on acceptable Web 2.0 practices and security features.

Besides that, the company shall adopt concrete IT policies to allow a safe inclusion of Web 2.0 technologies in the enterprise environment. A security solution that provides complete content protection, including application detection, monitoring and control is needed to discover threats embedded in Internet-based application traffic, and also to protect against data loss resulting from inappropriate use of social media applications. In addition, the content-based security enforcement is essential to mitigate these threats when they are discovered and to provide compete protection and threat elimination. Finally, other IT initiatives can be adopted such as high customized browse settings, installation of anti-malware software, adoption of strong authentication mechanisms and establishment of a data loss protection solution.

References

[1] G. Cormode and B. Krishnamurthy, "Key differences between Web 1.0 and Web 2.0", 2008, available on http://www2.research.att.com/~bala/papers/web1v2.pdf.
[2] A. Lipsman, "The network effect: facebook, linkedin, twitter & tumblr reach new heights in May", 2011, available on http://blog.comscore.com/2011/06/facebook_linkedin_twitter_tumblr.html.
[3] J. Sena, "The impact of Web 2.0 on technology", International Journal of Computer Science and Network Security, Vol. 9, No. 2, pp. 378-385, 2009.
[4] M. Chui, A. Miller and R. Roberts, "Six ways to make Web 2.0 work", McKinsey Quarterly, Vol. 1, pp. 2-7, 2009.
[5] O. Mahmood and S. Selvadurai, "Modelling web of trust with web 2.0", World Academy of Science, Engineering and Technology, Vol. 18, pp. 122-127, 2006.
[6] A. Jaokan and C. Sharma, "Mobile VoIP – approaching the tripping point", 2010, available on http://www.futuretext.com/downloads/mobile_voip.pdf.
[7] L. Kisselburgh, E. Spafford and M. Vorvoreanu, "Web 2.0: a complex balancing act", McAfee White Papers, pp. 2-13, 2010.
[8] J. Bughin and M. Chui, "The rise of the networked enterprise: Web 2.0 finds its payday", McKinsey Quarterly, Vol. IV, pp. 3-8, 2010.
[9] D. Rowe and C. Drew, "The impact of Web 2.0 on enterprise strategy", Journal of Information Technology Management, Vol. 19, No. 10, pp. 6-13, 2006.
[10] K. Yukihiro, "In-house use of Web 2.0: entrprise 2.0", NEC Technical Journal, Vol. 2, No. 2, pp. 46-49, 2007.
[11] S. Wright and J. Zdinak, "New communication behaviours in a Web 2.0 world", Alcatel Lucent White Papers, pp. 5-23, 2009.
[12] H. Mjemanze and D, Skeeles, "Expert guide to Web 2.0 threats: how to prevent an attack", ArcSight White Papers, pp. 3-14, 2010.
[13] G.Cluley and C. Theriault, "Security threat report: a look at the challenges ahead", Sophos White Papers, pp. 2-13, 2009.
[14] S. Pruitt, "Firewalls, the future and Web 2.0", 2007, available on http://www.networksecurityjournal.com/features/firewalls-the-future-web-2-061107/.
[15] G. Rydstedt, E. Bursztein, D. Boneh and C. Jackson, "Busting frame busting: a study of clickjacking vulnerabilities on popular sites", Proceedings of the IEEE Symposium on Security and Privacy, pp. 2-13, 2010.
[16] B. Wedge, "Data loss prevention: keeping your sensitive data out of the public domain", Insights on IT Risk Magazine, Vol. 9, pp. 2-23, 2011.
[17] S. Al-Fedaghi, "A conceptual foundation for data loss prevention", International Journal of Digital Content Technology and its Applications, Vol. 5, No. 3, pp. 293-303, 2011.

Authors Profile

Fernando Almeida is a pos-doc researcher at INESC TEC. He received the PhD. in Computer Engineering in 2010 from Faculty of Engineering of University of Porto. His research interests are in information security, Web 2.0 business models and enterprise architectures.